\documentclass[letterpaper,11pt]{article}

\usepackage{times}

\usepackage{amsthm,amsmath,amsfonts}

\usepackage{graphicx}
\usepackage[small]{caption}
\usepackage[hidelinks]{hyperref}
\newcommand{\orcid}[1]{\,\href{https://orcid.org/#1}{\includegraphics[width=8pt]{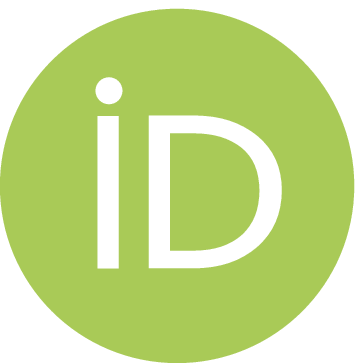}}}

\usepackage{color}

\renewcommand{\S}{\mathcal{S}}

\newcommand{\conv}{\mathrm{conv}}

\newtheorem{proposition}{Proposition}
\theoremstyle{remark}

\title{Disjoint axis-parallel segments without a circumscribing polygon}

\author{Rain Jiang\orcid{0000-0002-0144-942X}\qquad
Kai Jiang\orcid{0000-0001-8165-0571}\qquad
Minghui Jiang\orcid{0000-0003-1843-9292}\,\thanks{\texttt{ dr.minghui.jiang at gmail.com}}\medskip\\
Home School, USA}

\date{}

\begin{document}

\maketitle

\begin{abstract}
	We construct a family of $17$ disjoint axis-parallel line segments in the plane
	that do not admit a circumscribing polygon.
\end{abstract}

\section{Introduction}

For any family $\S$ of closed segments in the plane,
denote by $V(\S)$ the set of endpoints of the segments in $\S$.
A simple polygon $P$ is a \emph{circumscribing polygon} of $\S$ if the vertex set of $P$
is $V(\S)$, and every segment in $\S$ is either an edge or an internal diagonal in $P$.

\begin{figure}[htbp]
	\centering\includegraphics{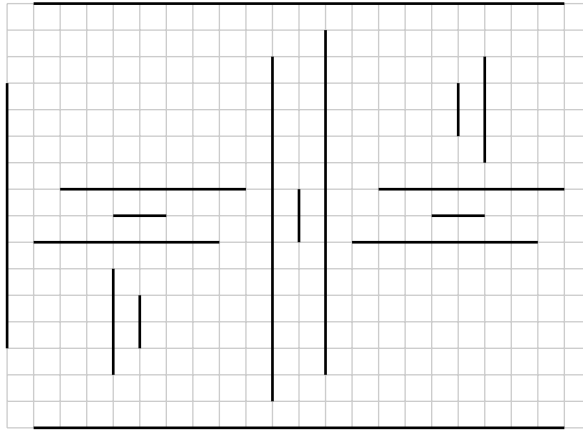}
	\caption{$17$ disjoint axis-parallel segments in a centrally symmetric configuration inside a $[-11,11]\times[-8,8]$ grid.}\label{fig:construction}
\end{figure}

Gr\"unbaum~\cite{Gr94} constructed a family $\S_4$
of six disjoint segments with four distinct slopes
that do not admit a circumscribing polygon.
Recently, Akitaya et~al.~\cite{AKRST19} constructed a family $\S_3$
of nine disjoint segments with three distinct slopes
that do not admit a circumscribing polygon,
and asked whether every family of disjoint axis-parallel segments in the plane,
not all in a line,
admit a circumscribing polygon.
In this note, we show that the family $\S_2$
of $17$ disjoint axis-parallel segments
illustrated in Figure~\ref{fig:construction}
do not admit a circumscribing polygon.

\section{The proof}

To prove that $\S_2$ does not admit a circumscribing polygon,
our main tool is the following proposition~\cite[Lemma~2.1]{BHT01}
which is repeatedly used by Akitaya et~al.~\cite{AKRST19} in proving
that the family $\S_3$ they constructed does not admit a circumscribing polygon.
For any polygon $P$,
denote by $\conv(P)$ the convex hull of $P$.

\begin{proposition}\label{prp}
	For any simple polygon $P$,
	the vertices in $\conv(P)$
	must appear in the same circular order in both $P$ and $\conv(P)$.
\end{proposition}

\begin{figure}[htbp]
	\centering\includegraphics{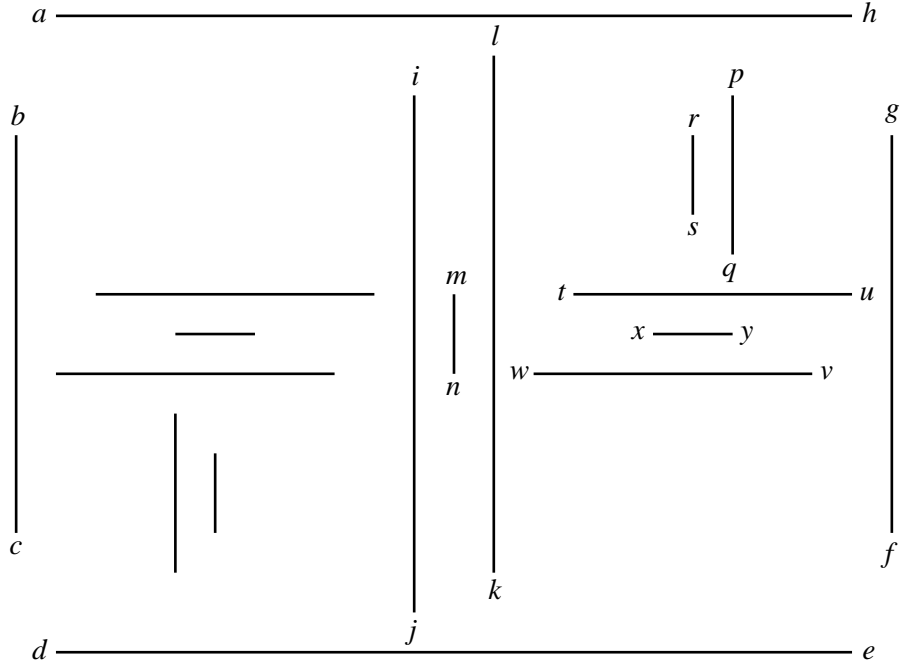}
	\caption{$17$ disjoint axis-parallel segments without a circumscribing polygon.}\label{fig:construction_a}
\end{figure}

Refer to Figure~\ref{fig:construction_a} for a magnified illustration of $\S_2$ with labels.
Suppose for contradiction that $\S_2$ admits a circumscribing polygon $P$.

Since the eight vertices $a,b,c,d,e,f,g,h$ are on the convex hull of $P$,
it follows by Proposition~\ref{prp} that
the four segments $ha, bc, de, fg$ must be edges in $P$.
Then $P$ is the alternating concatenation of these four edges
and four paths $a\to b, c\to d, e\to f, g\to h$.

We say that a path \emph{visits} a segment
if it goes through at least one endpoint of the segment.
Since the segments in $\S$ are in a centrally symmetric configuration,
we can assume without loss of generality that at least one of the two paths $a\to b$
and $c\to d$ visits the segment $mn$.
In the following we assume that this path is $a\to b$.
The other case, that this path is $c\to d$, is similar.

Let $P_{ab}$ be the simple polygon obtained by closing
the path $a \to b$ with the edge $ba$.
Suppose that $a\to b$ does not visit any segment to the right of $mn$.
Then it must visit both endpoints of $ij$.
Note that $a,b,i,j$ and $\{m,n\}\cap V(P_{ab})$
are all on $\conv(P_{ab})$.
Thus it follows by Proposition~\ref{prp}
that $ij$ must be an internal diagonal of $P_{ab}$ and hence an external diagonal of $P$.
Since $ij$ cannot be an external diagonal of the circumscribing polygon $P$,
$a\to b$ must visit at least one other segment to the right of $mn$.
Indeed, due to the strategic position of $kl$,
$a\to b$ must visit either $k$ or $l$ or both.

We claim that $a\to b$ must visit $l$.
Suppose the contrary.
Then $a\to b$ must visit $i,j,k$,
and cannot visit any segment to the right of $kl$.
Then $a,b,j,k,i$ are on $\conv(P_{ab})$,
and it again follows by Proposition~\ref{prp} that
$ij$ is external, a contradiction.

We have shown that $a\to b$ visits $l$.
We claim that $a\to b$ must not visit any segment to the right of $kl$.
Suppose the contrary,
and let $z\in\{ p,q,r,s,t,u,v,w,x,y \}$ be the rightmost endpoint that $a\to b$ visits,
breaking ties arbitrarily.
Then $a \to b$ must visit both $k$ and $l$,
and $a,b,k,z,l$ are on $\conv(P_{ab})$.
Then by Proposition~\ref{prp},
$kl$ is external, a contradiction.
Also, since $a\to b$ visits $l$,
$c\to d$ cannot visit any segment to the right of $kl$ either.

In summary,
the endpoints $\{ p,q,r,s,t,u,v,w,x,y \}$ must be visited by $e\to f$ and $g\to h$.
In addition, the endpoint $k$ may or may not be visited by $e\to f$ and $g\to h$.
In any case, since $l$ is visited by $a\to b$ or $c\to d$,
$e\to f$ and $g\to h$ cannot visit any segment to the left of $kl$.

We claim that $g\to h$ does not visit $xy$.
Suppose the contrary.
Let $z\in\{ x,y,w,v,k \}$ be the lowest endpoint that $g\to h$ visits.
Let $P_{gh}$ be the simple polygon obtained by closing
the path $g \to h$ with the edge $hg$.
Then $g\to h$ must visit both endpoints $u$ and $t$,
and $g,u,z,t,h$ are all on $\conv(P_{gh})$.
Then by Proposition~\ref{prp},
$ut$ is external, a contradiction.

Since $g\to h$ does not visit $xy$, $e\to f$ must visit $xy$.
Let $P_{ef}$ be the simple polygon obtained by closing
the path $e \to f$ with the edge $fe$.

We claim that $e\to f$ must visit $u$.
Suppose the contrary.
Then $e \to f$ may still visit $t$ but not any segment above $tu$.
Let $z\in\{ x,y,t \}$ be the highest endpoint that $e\to f$ visits.
Then $e\to f$ must visit both endpoints $w$ and $v$,
and $e,w,z,v,f$ are all on $\conv(P_{ef})$.
Then by Proposition~\ref{prp},
$wv$ is external, a contradiction.

We have shown that $e\to f$ visits $u$.
We claim that $e\to f$ must not visit any segment above $u$.
Suppose the contrary,
and let $z\in\{ p,q,r,s \}$ be the highest endpoint that $e\to f$ visits.
Then $e \to f$ must visit both $t$ and $u$,
and $e,t,z,u,f$ are on $\conv(P_{ef})$.
Then by Proposition~\ref{prp},
$tu$ is external, a contradiction.

In summary, the endpoints $\{ p,q,r,s \}$ must be visited by $g\to h$.
In addition, the endpoint $t$ may or may not be visited by $g\to h$.

Finally,
let $z\in\{ r,s,t \}$ be the leftmost endpoint that $g\to h$ visits.
Then $g \to h$ must visit both $p$ and $q$,
and $g,q,z,p,h$ are on $\conv(P_{gh})$.
Then by Proposition~\ref{prp},
$pq$ is external, a contradiction.

Therefore, our initial assumption that $\S_2$ admits a circumscribing polygon $P$
does not hold.
The proof is now complete.

\section{An open question}

Akitaya et~al.~\cite{AKRST19} proved that it is NP-hard to decide
whether a given family of disjoint segments admit a circumscribing polygon.
Is this decision problem still NP-hard on disjoint axis-parallel segments?

\end{document}